\documentstyle[12pt]{article}
\begin{document}
\hfill{NCKU-HEP-98-15}\par
%\hfill{hep-ph/9812XXX}
\vskip 0.3cm
\begin{center}
{\large {\bf Unification of the $k_T$ and threshold resummations}}
\vskip 1.0cm
Hsiang-nan Li
\vskip 0.3cm
Department of Physics, National Cheng-Kung University, \par
Tainan 701, Taiwan, Republic of China
\end{center}
\vskip 1.0cm

%PACS numbers: 12.38.Bx, 12.38.Cy
\vskip 1.0cm
%\baselineskip=2\baselineskip

\centerline{\bf Abstract}
\vskip 0.3cm

We derive a resummation formula for a $k_T$-dependent parton distribution
function at threshold, where $k_T$ is a parton transverse momentum. The
derivation requires infrared cutoffs for both longitudinal and transverse
loop momenta as evaluating soft gluon emissions in the Collins-Soper
resummation framework. This unified resummation exhibits suppression at
large $b$, $b$ being the conjugate variable of $k_T$, which is similar to
the $k_T$ resummation, and exhibits enhancement at small $b$, similar to
the threshold resummation.

\newpage

\centerline{\large \bf 1. Introduction}
\vskip 0.5cm

Recently, we have demonstrated \cite{L9} that both the $k_T$ resummation
for a parton distribution function $\phi(x,k_T,p^+)$ and the threshold
resummation for $\phi(x,p^+)$ can be performed in the Collins-Soper (CS)
framework \cite{CS}. The distribution function $\phi(x,k_T,p^+)$,
associated with a hadron of momentum $p^\mu=p^+\delta^{\mu+}$, describes
the probability that a parton carries the longitudinal momentum $xp^+$ and
the transverse momentum $k_T$. The distribution function $\phi(x,p^+)$,
which coincides with the standard parton model, comes from 
$\phi(x,k_T,p^+)$ with $k_T$ integrated out. 
In the $k_T$ resummation the double logarithms $\ln^2(p^+b)$, $b$ being 
the conjugate variable of $k_T$, are organized. The result is Sudakov 
suppression, quoted as \cite{L1}
\begin{equation}
\phi(x,b,p^+)=\exp\left[-2\int_{1/b}^{x p^+}\frac{d p}{p}
\int_{1/b}^{p}\frac{d\mu}{\mu}
\gamma_{K}(\alpha_s(\mu))\right]\phi^{(0)}\;,
\label{fb0}
\end{equation}
where the anomalous dimension $\gamma_K$ will be defined later, and
$\phi^{(0)}$ is the initial condition of the double-logarithm evolution. 
In the threshold resummation the double logarithms $\ln^2(1/N)$, $N$
being the moment of $\phi(x,p^+)$, are organized. The result is an 
enhancement \cite{S}:
\begin{eqnarray}
\phi(N,p^+)=\exp\left[-2\int_{0}^{1}dz\frac{z^{N-1}-1}{1-z}
\int_{1-z}^{1}\frac{d\lambda}{\lambda}
\gamma_{K}(\alpha_s(\lambda p^+))\right]\phi^{(0)}\;,
\label{ft2}
\end{eqnarray}
with the same anomalous dimension $\gamma_K$.

We have neglected the dependence of $\phi$ on the renormalization (or 
factorization) scale $\mu$, which denotes the single-logarithm evolution, 
and can be easily derived using renormalization-group (RG) equations.
Equation (\ref{fb0}) has been employed to evaluate the $p_T$ spectrum of 
direct photon production, $p_T$ being the transverse momentum of the direct 
photon \cite{LL}. It was observed that the Sudakov exponential, 
after Fourier transformed to the $k_T$ space, provides the necessary $k_T$ 
smearing effect, which resolves the discrepancy between experimental data
and the next-to-leading-order QCD ($\alpha\alpha_s^2$) predictions
\cite{HKK}. Equation (\ref{ft2}) is suitable for the analyses of dijet,
direct photon and heavy quark productions in kinematic end-point regions
\cite{S}.

It has been found that in the CS framework the two different types of
double logarithms $\ln^2(p^+b)$ and $\ln^2(1/N)$ are summed by
choosing appropriate infrared cutoffs in the evaluation of soft gluon
corrections. If transverse degrees of freedom of a parton are included,
$1/b$ will serve as an infrared cutoff for soft gluon emissions. Combined
with the scale $p^+$ from the hadron momentum, the large double logarithms
$\ln^2(p^+b)$ are generated. In the end-point region with $x\to 1$, we keep
the longitudinal cutoff $(1-x)p^+$, or $p^+/N$ in the moment space, and
integrate out the transverse degrees of freedom of a parton. Combined with
$p^+$, the double logarithms $\ln^2(1/N)$ are produced. Therefore, in
the case with $p^+b\gg N$, we neglect the longitudinal cutoff, and
sum $\ln^2(p^+b)$. In the case with $N\gg p^+b$, we neglect the
transverse cutoff, and sum $\ln^2(1/N)$. Based on these observations, it is
straightforward to develop a unified resummation formalism by retaining
the longitudinal and transverse cutoffs simultaneously. It will be shown
in this letter
that this unified resummation exhibits suppression at large $b$, similar to
the $k_T$ resummation, and exhibits enhancement at small $b$, similar to
the threshold resummation.

\vskip 1.0cm
\centerline{\large \bf 2. Formalism}
\vskip 0.5cm

Consider a quark distribution function for a hadron in the minimal
subtraction scheme,
\begin{equation}
\phi(x,k_T,p^+)=\int\frac{dy^-}{2\pi}\int\frac{d^2 b}{(2\pi)^2}
e^{-ix p^+y^-+i{\bf k}_T\cdot {\bf b}}
\langle p| {\bar q}(y^-,{\bf b})\frac{1}{2}\gamma^+q(0)|p\rangle\;,
\label{dep}
\end{equation}
where $\gamma^+$ is a Dirac matrix, and $|p\rangle$ denotes the
hadron with the momentum $p$. Averages over spin
and color are understood. The above definition is given in the axial gauge
$n\cdot A=0$, where the gauge vector $n$ is assumed to be arbitrary with
$n^2\not= 0$. Though this definition is gauge dependent, physical
observables, such as hadron structure functions and cross sections, are
gauge invariant. It has been shown that the $n$ dependences cancel among
the convolution factors, {\it i.e.}, among parton distribution functions,
final-state jets, and nonfactorizable soft gluon exchanges, in the
factorization formula for a DIS structure function \cite{L1}.

The key step in the CS technique is to obtain the derivative
$p^+d\phi/dp^+$ \cite{CS}. In the axial gauge $n$ appears in the
gluon propagator, $(-i/l^2)N^{\mu\nu}(l)$, with
\begin{equation}
N^{\mu\nu}(l)=g^{\mu\nu}-\frac{n^\mu l^\nu+n^\nu l^\mu}
{n\cdot l}+n^2\frac{l^\mu l^\nu}{(n\cdot l)^2}\;.
\label{gp}
\end{equation}
Because of the scale invariance of $N^{\mu\nu}$ in $n$, $\phi$ depends on
$p^+$ through the ratio $(p\cdot n)^2/n^2$, implying that the differential
operator $d/dp^+$ can be replaced by $d/dn_\alpha$ using a chain rule,
\begin{equation}
p^+\frac{d}{dp^+}\phi=-\frac{n^2}{v\cdot n}v_{\alpha}
\frac{d}{dn_\alpha}{\phi}\;,
\label{cr}
\end{equation}
where $v=(1,0,{\bf 0})$ is a dimensionless vector along $p$.
The operator $d/dn_\alpha$ applies to $N^{\mu\nu}$, leading to
\begin{eqnarray}
-\frac{n^2}{v\cdot n}v_{\alpha}
\frac{d}{dn_\alpha}N^{\mu\nu}= {\hat v}_{\alpha}
\left(N^{\mu\alpha}l^\nu+N^{\alpha\nu}l^\mu\right)\;,
\label{dgp}
\end{eqnarray}
with the special vertex
\begin{equation}
{\hat v}_{\alpha}=\frac{n^2v_{\alpha}}{v\cdot nn\cdot l}\;.
\label{sve}
\end{equation}

The momentum $l^\mu$ ($l^\nu$) is contracted with the vertex the
differentiated gluon attaches, which is then replaced by the special vertex
in Eq.~(\ref{sve}). For each type of vertices, there exists a Ward identity,
which relates the diagram with the contraction of $l^\mu$ ($l^\nu$) to the
difference of two diagrams \cite{L3}. A pair cancellation occurs between
the contractions with two adjacent vertices. Summing the diagrams
containing different differentiated gluons, the special vertex moves to the
outer end of a parton line. We arrive at the derivative,
\begin{equation}
p^+\frac{d}{dp^+}\phi(x,k_T,p^+)=2{\bar \phi}(x,k_T,p^+)\;,
\label{dif}
\end{equation}
described by Fig.~1(a), where the square in the new function
${\bar \phi}$ represents the special vertex ${\hat v}_\alpha$. The
coefficient 2 comes from the equality of the two new functions with
the special vertex on either side of the final-state cut.

To obtain a differential equation of $\phi$ from Eq.~(\ref{dif}), we need
to factorize the subdiagram containing the special vertex out of $\bar\phi$.
The factorization holds in the leading regions of the loop momentum $l$ that
flows through the special vertex. The collinear region of $l$ is not leading
because of the factor $1/(n\cdot l)$ in ${\hat v}_\alpha$ with nonvanishing
$n^2$. Therefore, the leading regions of $l$ are soft and hard, in which the
subdiagram is factorized from ${\bar \phi}$ into a soft function $K$ and a
hard function $G$, respectively. The remaining part is the original
distribution function $\phi$. That is, $\bar\phi$ is expressed as the
convolution of the functions $K$ and $G$ with $\phi$.

The lowest-order contribution to $K$ is extracted from Fig.~1(b), whose
contribution is written as
\begin{equation}
{\bar \phi}_s(x,k_T,p^+)={\bar \phi}_{sv}(x,k_T,p^+)+
{\bar \phi}_{sr}(x,k_T,p^+)\;,
\label{fss}
\end{equation}
with
\begin{eqnarray}
{\bar \phi}_{sv}&=&\left[ig^2C_F\mu^\epsilon
\int\frac{d^{4-\epsilon}l}{(2\pi)^{4-\epsilon}}
N_{\nu\beta}(l)\frac{{\hat v}^\beta v^\nu}{v\cdot l}
\frac{1}{l^2}-\delta K\right]\phi(x,k_T,p^+)\;,
\label{fsv} \\
{\bar \phi}_{sr}&=&ig^2C_F\mu^\epsilon
\int\frac{d^{4-\epsilon}l}{(2\pi)^{4-\epsilon}}
N_{\nu\beta}(l)\frac{{\hat v}^\beta v^\nu}{v\cdot l}
2\pi i\delta(l^2)
\nonumber\\
& &\times\phi(x+l^+/p^+,|{\bf k}_T+{\bf l}_T|,p^+)\;,
\label{fsr}
\end{eqnarray}
corresponding to the virtual and real gluon emissions, respectively.
$C_F=4/3$ is a color factor, and $\delta K$ an additive counterterm.
The ultraviolet pole in Eq.~(\ref{fsv}) is isolated using the dimensional
regularization.

To work out the $l_T$ integration explicitly, we employ the the Fourier
transform from the $k_T$ space to the $b$ space. Inserting the identities
\begin{eqnarray}
\int_x^1d\xi \delta(\xi-x)=1\;,\;\;\;\;
\int_x^1d\xi \delta(\xi-x-l^+/p^+)=1\;,
\end{eqnarray}
into Eqs.~(\ref{fsv}) and (\ref{fsr}), respectively, Eq.(\ref{fss}) becomes
\begin{eqnarray}
{\bar \phi}_s(x,b,p^+)=\int_x^1\frac{d\xi}{\xi}
K\left(\left(1-\frac{x}{\xi}\right)\frac{p^+}{\mu},b\mu,
\alpha_s(\mu)\right)\phi(\xi,b,p^+)\;,
\label{tss1}
\end{eqnarray}
with
\begin{eqnarray}
K&=&ig^2C_F\mu^\epsilon\int\frac{d^{4-\epsilon}l}{(2\pi)^{4-\epsilon}}
N_{\nu\beta}(l)\frac{{\hat v}^\beta v^\nu}{v\cdot l}
\left[\frac{\delta(1-x/\xi)}{l^2}\right.
\nonumber\\
& &\left.+2\pi i\delta(l^2)\delta\left(1-\frac{x}{\xi}-\frac{l^+}{p^+}
\right)e^{i{\bf l}_T\cdot {\bf b}}\right]
-\delta K\delta\left(1-\frac{x}{\xi}\right)\;.
\label{kss}
\end{eqnarray}
To obtain the above expression, we have adopted the approximation
\begin{eqnarray}
\delta\left(1-\frac{x}{\xi}-\frac{l^+}{\xi p^+}\right)
\approx\delta\left(1-\frac{x}{\xi}-\frac{l^+}{p^+}\right)\;,
\end{eqnarray}
which is appropriate in the threshold region with $x\to 1$. The exponential
$\exp(i{\bf l}_T\cdot {\bf b})$ in the second term arises form the Fourier
transform of the real gluon contribution.

To work out the $\xi$ integration explicitly, we further employ the
Mellin transform from the momentum fraction ($x$) space to the moment ($N$)
space:
\begin{eqnarray}
{\bar\phi}_{s}(N,b,p^+)&\equiv &\int_0^1 dxx^{N-1}{\bar\phi}_{s}(x,b,p^+)\;,
\nonumber\\
&=&K(p^+/(N\mu),b\mu,\alpha_s(\mu))\phi(N,b,p^+)\;.
\end{eqnarray}
with
\begin{eqnarray}
K(p^+/(N\mu),b\mu,\alpha_s(\mu))=\int_0^1 dzz^{N-1}
K((1-z)p^+/\mu,b\mu,\alpha_s(\mu))\;.
\label{kmt}
\end{eqnarray}
The convolutions between $K$ and $\phi$ in $l^+$ and $l_T$ are then
completely simplified into a multiplication under the Mellin and Fourier
transforms, respectively.

Equation (\ref{kmt}) is evaluated in the Appendix, and the result is
\begin{eqnarray}
K(p^+/(N\mu),b\mu,\alpha_s(\mu))
=\frac{\alpha_s(\mu)}{\pi}C_F\left[\ln\frac{1}{b\mu}
-K_0\left(\frac{2\nu p^+b}{N}\right)\right]\;,
\label{uk}
\end{eqnarray}
$K_0$ being the modified Bessel function. The gauge factor 
$\nu=\sqrt{(v\cdot n)^2/|n^2|}$ confirms our argument that $\phi$
depends on $p^+$ via the ratio $(p\cdot n)^2/n^2$. It is easy to examine 
the large $p^+b$ and $N$ limits of the above expression. For $p^+b\gg N$, 
we have $K_0\to 0$ and 
\begin{equation}
K\to\frac{\alpha_s}{\pi}C_F\ln\frac{1}{b\mu}\;,
\label{lpb}
\end{equation}
which is exactly the soft function with the characteristic scale $1/b$
appearing in the $k_T$ resummation \cite{L9,L1}. For $N\gg p^+b$, we have
$K_0\approx -\ln(\nu p^+b/N)$ and 
\begin{equation}
K\to\frac{\alpha_s}{\pi}C_F\ln\frac{\nu p^+}{N\mu}\;,
\label{ln}
\end{equation}
which is the soft function with the characteristic scale $p^+/N$ for the
threshold resummation \cite{L9}. Hence, Eq.~(\ref{uk}) is indeed appropriate
for the unification of the $k_T$ and threshold resummations.

The lowest-order contribution to $G$ from Fig.~1(c) is given by
\begin{equation}
{\bar\phi}_h(N,b,p^+)=G(p^+/\mu,\alpha_s(\mu))\phi(N,b,p^+)\;,
\label{ght}
\end{equation}
in the $b$ and $N$ spaces, with
\begin{eqnarray}
G=-ig^2C_F\mu^\epsilon\int\frac{d^{4-\epsilon}l}{(2\pi)^{4-\epsilon}}
N_{\nu\beta}(l)\frac{{\hat v}^\beta}{l^2}
\left[\frac{\not p-\not l}{(p- l)^2}\gamma^\nu
+\frac{v^\nu}{v\cdot l}\right]
-\delta G\;,
\label{gpb}
\end{eqnarray}
in the limit $x\to 1$, where $\delta G$ is an additive counterterm. The
second term, whose sign is opposite to that of ${\bar\phi}_{sv}$, is a
soft subtraction. This term avoids double counting, and ensures a hard
momentum flow in $G$. A straightforward calculation gives
\begin{eqnarray}
G(p^+/\mu,\alpha_s(\mu))=-\frac{\alpha_s(\mu)}{\pi}C_F
\ln\frac{p^+\nu}{\mu}\;,
\label{gh}
\end{eqnarray}
which is the same as the hard functions with the characteristic scale $p^+$
appearing in the $k_T$ and threshold resummations \cite{L9,L1}. 

Using ${\bar\phi}={\bar\phi}_s+{\bar\phi}_h$, Eq.~(\ref{dif}) becomes
\begin{eqnarray}
p^+\frac{d}{dp^+}\phi(N,b,p^+)&=&2\left[K(p^+/(N\mu),b\mu,\alpha_s(\mu))+
G(p^+/\mu,\alpha_s(\mu))\right]
\nonumber\\
& &\times\phi(N,b,p^+)\;.
\label{dph}
\end{eqnarray}
The functions $K$ and $G$ possess ultraviolet divergences individually as
indicated by their counterterms. These divergences, both from the virtual
gluon contribution ${\bar \phi}_{sv}$, cancel each other, such that the sum
$K+G$ is RG invariant. The single logarithms contained in $K$ and $G$
are organized by the RG equations
\begin{equation}
\mu\frac{d}{d\mu}K=-\gamma_K=-\mu\frac{d}{d\mu}G\;.
\label{kg}
\end{equation}
The anomalous dimension of $K$, $\lambda_K=\mu d\delta K/d\mu$,
is given, up to two loops, by \cite{BS}
\begin{equation}
\gamma_K=\frac{\alpha_s}{\pi}C_F+\left(\frac{\alpha_s}{\pi}
\right)^2C_F\left[C_A\left(\frac{67}{36}
-\frac{\pi^{2}}{12}\right)-\frac{5}{18}n_{f}\right]\;,
\label{lk}
\end{equation}
with $n_{f}$ the number of quark flavors, and $C_A=3$ a color factor.

As solving Eq.~(\ref{kg}), we allow the variable $\mu$ evolves from the
characteristic scale of $K$ to the scale of $G$. We discuss the
cases for $p^+b\gg N$ and for $N\gg p^+b$ first, which will help the
derivation of the unified resummation. The solution of $K+G$ is written as 
\begin{eqnarray}
& &K(p^+/(N\mu),b\mu,\alpha_s(\mu))+G(p^+/\mu,\alpha_s(\mu))
\nonumber\\
&=&-\int_{1/b}^{p^+}\frac{d\mu}{\mu}\gamma_K(\alpha_s(\mu))\;,
\;\;\;\;{\rm for}\;\;p^+b\gg N\;,
\nonumber\\
& &-\int_{p^+/N}^{p^+}\frac{d\mu}{\mu}\gamma_K(\alpha_s(\mu))\;,
\;\;\;\;{\rm for}\;\;N\gg p^+b\;,
\label{skg}
\end{eqnarray}
where the initial conditions of $K$ and $G$ of the RG evolution have been
neglected, since they are irrelevant to the double-logarithm summation.
Equation (\ref{skg}) indicates that the distribution function
$\phi(N,b,p^+)$ involves $\ln(p^+b)$ and $\ln(1/N)$ in
the $p^+b\to \infty$ and $N\to \infty$ limits, respectively, as stated
before. This can be easily understood by ignoring the variation of
$\gamma_K$, and performing the $\mu$ integration directly.

Inserting Eq.~(\ref{skg}) into (\ref{dph}), we obtain the solutions 
\begin{eqnarray}
\phi(N,b,p^+)&=&\exp\left[-2\int_{1/b}^{p^+}\frac{d p}{p}
\int_{1/b}^{p}\frac{d\mu}{\mu}
\gamma_{K}(\alpha_s(\mu))\right]\phi^{(0)}\;,
\label{fb1}\\
\phi(N,b,p^+)&=&\exp\left[-2\int_{p^+}^{p^+/N}\frac{dp}{p}
\int_{p}^{p^+}\frac{d\mu}{\mu}
\gamma_{K}(\alpha_s(\mu))\right]\phi^{(0)}\;,
\label{ft3}
\end{eqnarray}
for $p^+b\gg N$ and $N\gg p^+b$, respectively. Note that we have replaced
the derivative $p^+d/dp^+$ by $(p^+/N)d/d(p^+/N)$ as deriving
Eq.~(\ref{ft3}) \cite{L9}, since we intend to resum 
$\ln(1/N)$. Obviously, Eq.~(\ref{fb1}) is identical to Eq.~(\ref{fb0}) with
$x\to 1$, and Eq.~(\ref{ft3}) is equivalent to Eq.~(\ref{ft2}) up to
corrections suppressed by a power $1/N$ \cite{L9}.

Equation (\ref{uk}) implies the characteristic scale of $K$ for the
unified resummation,
\begin{equation}
\frac{1}{b}\exp\left[-K_0\left(\frac{p^+b}{N}\right)\right]\;,
\label{usc}
\end{equation}
where the gauge factor $2\nu$ has been suppressed.
Hinted by Eqs.~(\ref{lpb}) and (\ref{ln}), we rewrite Eqs.~(\ref{fb1}) and
(\ref{ft3}) in terms of the above scale as
\begin{eqnarray}
\phi(N,b,p^+)&=&\exp\left[-2\int_{\exp[-K_0(p^+b/N)]/b}^{p^+}\frac{d p}{p}
\int_{1/b}^{p}\frac{d\mu}{\mu}
\gamma_{K}(\alpha_s(\mu))\right]\phi^{(0)}\;,
\label{fb2}\\
\phi(N,b,p^+)&=&\exp\left[-2\int_{\exp[-K_0(p^+b/N)]/b}^{p^+}\frac{dp}{p}
\int_{p^+}^{p}\frac{d\mu}{\mu}
\gamma_{K}(\alpha_s(\mu))\right]\phi^{(0)}\;,
\label{ft4}
\end{eqnarray}
At last, to unify the above expressions, we replace the lower bounds
of $\mu$ by
\begin{equation}
\frac{1}{b}\exp[-K_0(p^+b)]\;,
\end{equation}
which is motivated by Eq.~(\ref{usc}). The unified resummation is then
given by
\begin{equation}
\phi(N,b,p^+)=\exp\left[-2\int_{\exp[-K_0(p^+b/N)]/b}^{p^+}\frac{d p}{p}
\int_{\exp[-K_0(p^+b)]/b}^{p}\frac{d\mu}{\mu}
\gamma_{K}(\alpha_s(\mu))\right]\phi^{(0)},
\label{fb3}
\end{equation}
which is appropriate for arbitrary $p^+b$ and $N$. It is easy to justify
that Eq.~(\ref{fb3}) approaches the $k_T$ resummation in Eq.~(\ref{fb1})
as $b\to\infty$, and approaches the threshold resummation in
Eq.~(\ref{ft3}) as $b\to 0$.

\vskip 1.0cm
\centerline{\large \bf 3. Discussion}
\vskip 0.5cm

Comparing Eq.~(\ref{fb3}) with (\ref{fb1}) and (\ref{ft3}), we have the
following remark. In the $k_T$ resummation it is the logarithms of the
large scale $p^+$ from $G$ that are organized, and the result is a
suppression. In the threshold resummation it is the logarithms of the small
scale $p^+/N$ from $K$, as shown in Eq.~(\ref{ln}), that are organized. The
result is an enhancement. The opposite effects of these two resummations
are attributed to the opposite directions of the double-logarithm evolution:
from $1/b$ to the large $p^+$ in the former case, and from $p^+$ to the
small $p^+/N$ in the latter case. The unified resummation for a
$k_T$-dependent parton distribution function at threshold exhibits both
behaviors: it is a suppression in the large $b$ region ($p^+b\gg N$), and
turns into an enhancement in the small $b$ region ($N\gg p^+b$). That is,
Eq.~(\ref{fb3}) displays the opposite effects of the $k_T$ and
threshold resummations at different $b$.

The behavior of the unified resummation can be explained as follows. For an
intermediate $x$, virtual and real soft gluon corrections cancel exactly
in the small $b$ region, since they have almost equal phase space. Hence,
there are only single collinear logarithms, namely, no double logarithms.
In this case the Sudakov exponential approaches unity as $b< 1/p^+$
\cite{LS}, indicating the soft cancellation stated above. However, at
threshold ($x\to 1$), real gluon emissions still do not have sufficient
phase space even as $b\to 0$, and soft virtual corrections are not
cancelled exactly. In this case the double logarithms $\ln^2(1/N)$ persist
and become dominant. Sudakov suppression then transits into an enhancement,
instead of unity, smoothly as $b$ decreases.

The result obtained in this work will be useful for the study of, say, the
production of large $p_T$ direct photons or jets. Phenomenological
applications of Eq.~(\ref{fb3}) will be published elsewhere.

\vskip 0.5cm
This work was supported by the National Science Council of Republic of
China under Grant No. NSC-88-2112-M-006-013.

\vskip 1.0cm
\centerline{\large\bf Appendix}
\vskip 0.5cm

In this Appendix we present the detail of deriving Eq.~(\ref{uk}). A simple
investigation of Eq.~(\ref{fss}) shows that the first term $g_{\nu\beta}$
in $N_{\nu\beta}$ gives a vanishing contribution because of $v^2=0$, and
the contribution from the second term $-n_\nu l_\beta/n\cdot l$ cancels the
contribution from the fourth term $n^2l_\nu l_\beta/(n\cdot l)^2$. Hence,
we concentrate only on the third term $-n_\beta l_\nu/n\cdot l$, which
leads Eq.~(\ref{kmt}) to
\begin{eqnarray}
K&=&-ig^2C_F\mu^\epsilon\int_0^1 dzz^{N-1}
\int\frac{d^{4-\epsilon}l}{(2\pi)^{4-\epsilon}}
\frac{n^2}{(n\cdot l)^2}
\left[\frac{\delta(1-z)}{l^2}\right.
\nonumber\\
& &\left.+2\pi i\delta(l^2)\delta\left(1-z-\frac{l^+}{p^+}
\right)e^{i{\bf l}_T\cdot {\bf b}}\right]
-\delta K\delta(1-z)\;.
\end{eqnarray}
Performing the integrations over $l^-$ and $l^+$, we have
\begin{eqnarray}
K&=&-g^2C_F\mu^\epsilon\int_0^1 dzz^{N-1}
\int\frac{d^{2-\epsilon}l}{(2\pi)^{3-\epsilon}}
\left[\frac{\delta(1-z)}{l_T^2}\right.
\nonumber\\
& &\left.-\frac{2n^2(1-z)p^{+2}}{[n^+l_T^2+2n^-(1-z)^2p^{+2}]^2}
e^{i{\bf l}_T\cdot {\bf b}}\right]
-\delta K\delta(1-z)\;.
\end{eqnarray}

The first term in the integral has been evaluated in \cite{L1}, and the
result with its ultraviolet pole subtracted by $\delta K$ is quoted as
\begin{equation}
\frac{\alpha_s}{\pi}C_F\ln \frac{a}{\mu}\;,
\label{tsv1}
\end{equation}
where constants of order unity have been neglected. The infrared regulator
$a$, introduced by the replacement of the denominator $l_T^2$ by
$l_T^2+a^2$, will approach zero at last. The second term, after performing
the integration over $l_T$, gives
\begin{eqnarray}
2\frac{\alpha_s}{\pi}C_F\nu p^+b\int_0^1 dzz^{N-1}
K_1\left(2\sqrt{(1-z)^2\nu^2 p^{+2}+a^2}b\right)\;,
\label{2nd}
\end{eqnarray}
where $K_1$ is the modified Bessel function, and the infrared regulator $a$
for the divergence at $z\to 1$ is also introduced by the same replacement.

We adopt the identity
\begin{eqnarray}
& &\int_0^1 dzz^{N-1}K_1\left(2\sqrt{(1-z)^2\nu^2 p^{+2}+a^2}b\right)
\nonumber\\
& &=\int_0^1 dz(z^{N-1}-1)K_1(2(1-z)\nu p^+b)
\nonumber\\
& &+\int_0^1 dz K_1\left(2\sqrt{(1-z)^2\nu^2 p^{+2}+a^2}b\right)\;,
\label{zi}
\end{eqnarray}
where $a$ in the first term has been dropped, since the $z$-integral is
infrared finite. To perform the integration, we employ the relation
\begin{eqnarray}
\int_0^1 dz(z^{N-1}-1)K_1(2(1-z)\nu p^+b)
=-\int_0^{1-1/N} dz K_1(2(1-z)\nu p^+b)\;,
\end{eqnarray}
which is is valid up to corrections suppressed by $1/N$. It is then
trivial to show that Eq.~(\ref{2nd}) leads to
\begin{eqnarray}
\frac{\alpha_s}{\pi}C_F\left[\ln\frac{1}{ab}
-K_0\left(\frac{2\nu p^+b}{N}\right)\right]\;.
\label{uk0}
\end{eqnarray}
At last, combining Eqs.~(\ref{tsv1}) and (\ref{uk0}), we obtain
Eq.~(\ref{uk}).

\newpage
\centerline{\large \bf Figure Captions}
\vskip 0.3cm

\noindent
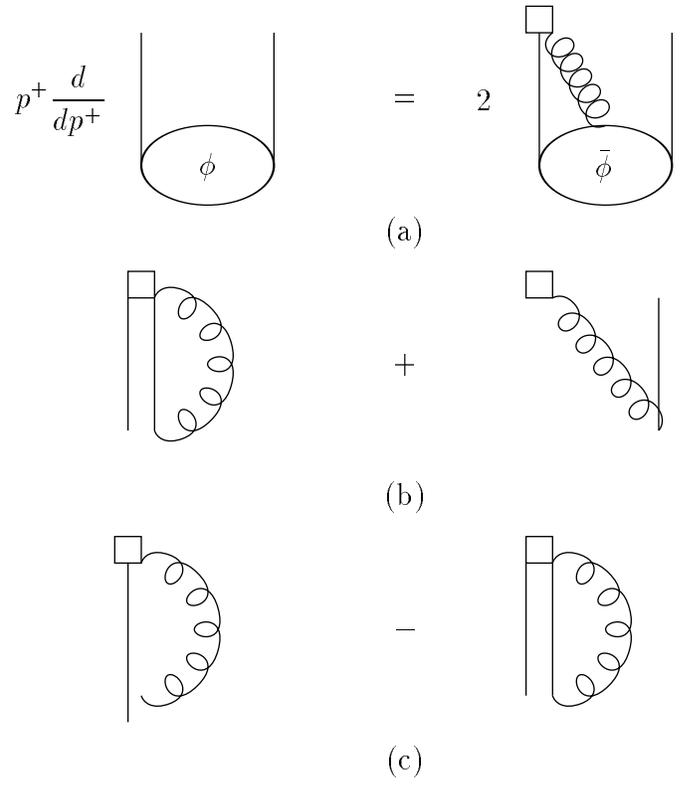
{\bf Fig. 1.} (a) The derivative $p^+d\phi/dp^+$ in the axial gauge. (b)
The $O(\alpha_s)$ function $K$. (c) The $O(\alpha_s)$ function $G$.
\vskip 0.3cm


\newpage
\begin{thebibliography}{99}
\bibitem{L9} H-n. Li, Report No. hep-ph/9811340.
\bibitem{CS} J.C. Collins and D.E. Soper, Nucl. Phys. {\bf B193} (1981) 381.
\bibitem{L1} H-n. Li, Phys. Lett. B {\bf 369} (1996) 137;
Phys. Rev. D {\bf 55} (1997) 105.
\bibitem{S} N. Kidonakis, G. Oderda, and G. Sterman, Report No.
hep-ph/9801268; E. Laenen, G. Oderda, and G. Sterman, Report No.
hep-ph/9806467.
\bibitem{LL} H.L. Lai and H-n. Li, to appear in Phys. Rev. D.
\bibitem{HKK} J. Huston {\it et al.}, Phys. Rev. D {\bf 51} (1995) 6139.
%\bibitem{LY} H-n. Li and H.L. Yu, Phys. Rev. D {\bf 53}, 4970 (1996);
%{\bf 55}, 2833 (1997).
\bibitem{L3} H-n. Li, Phys. Lett. B {\bf 405} (1997) 347; Report No.
hep-ph/9807392, to appear in Chin. J. Phys.
\bibitem{BS} J. Botts and G. Sterman, Nucl. Phys. {\bf B325} (1989) 62.
\bibitem{LS} H-n. Li and G. Sterman, Nucl. Phys. {\bf B381} (1992) 129.


\end{thebibliography}
\end{document}